\documentclass[12pt,twoside]{article}
\usepackage{xrb2007}
\usepackage{natbib,graphics,graphicx,subfigure}

\newcommand{\eg}{{\sl e.g.~}}

\newcommand{\etal}{{\sl et al.}}

\newcommand{\JHK}{$J\!H\!K~$}
\newcommand{\degs}{$^{\circ}~$}

\newcommand{\ergs}{$\,$erg$\,$s$^{-1}$}
\newcommand{\Chandra}{{\it Chandra~}}

\begin{document}

\session{Faint Galactic XRB Populations}

\shortauthor{Bandyopadhyay \etal}
\shorttitle{The Path to Buried Treasure}

\title{The Path to Buried Treasure: Paving the Way to the FLAMINGOS-2 Galactic Center Survey with IR and X-ray Observations}
\vspace{-2mm}
\author{Reba M. Bandyopadhyay, Stephen S. Eikenberry, Curtis Dewitt}
\affil{Dept. of Astronomy, University of Florida, Gainesville, FL 32611}
\author{Andrew J. Gosling}
\affil{Dept. of Astrophysics, University of Oxford, Oxford OX1 3RH, UK}
\author{Michael P. Muno}
\affil{California Institute of Technology, Pasadena, CA 91125}
\author{and the F2GCS Team}

\begin{abstract}
I describe the IR and X-ray campaign we have undertaken to determine
the nature of the faint discrete X-ray source population discovered by
\Chandra in the Galactic Center.  These results will provide the input
to the FLAMINGOS-2 Galactic Center Survey (F2GCS).  With FLAMINGOS-2's
multi-object IR spectrograph we will obtain 1000s of IR spectra of
candidate X-ray source counterparts, allowing us to efficiently
identify the nature of these sources, and thus dramatically increase
the number of known X-ray binaries and CVs in the Milky Way.
\end{abstract}

\vspace{-8mm}
\section{Introduction}
\vspace{-1mm} 
The unprecedented sensitivity and angular resolution of \Chandra has
been utilized by Wang \etal~ (2002; hereafter W02) and Muno \etal~
(2003; hereafter M03) to investigate the X-ray source population of
the Galactic Center (GC).  The W02 ACIS-I survey of the central
0.8\degs$\times$2\degs of the GC revealed a large population of
previously undiscovered discrete weak sources with X-ray luminosities
of $10^{32}-10^{35}$\ergs.  M03 imaged the central 40 pc$^{2}$ (at 8
kpc) around Sgr A* (Fig. 1a), finding an additional $\sim$2300
discrete point sources down to a limiting flux of $10^{31}$ erg/s.
Some individual sources have been identified as X-ray transients,
high-mass stars, LMXBs, and CVs.  However, the nature of the majority
of these newly detected sources is as yet unknown.

We have undertaken an IR and X-ray observational campaign to determine
the nature of the faint discrete X-ray source population discovered by
\Chandra in the GC.  Data obtained to date includes a deep \Chandra
survey of the Nuclear Bulge; deep, high resolution IR imaging from
VLT/ISAAC, CTIO/ISPI, and the UKIDSS Galactic Plane Survey; and IR
spectroscopy from VLT/ISAAC and IRTF/SpeX.  By cross-correlating the
X-ray imaging with our IR surveys, we initially identify candidate
counterparts to the X-ray sources via astrometry.  Using a detailed IR
extinction map, we are deriving magnitudes and colours for all the
candidates.  Having thus established a target list, we will use the
multi-object IR spectrograph FLAMINGOS-2 on Gemini South to carry out
a spectroscopic survey of the candidates, to search for the emission
lines which are a hallmark of accreting binaries.  By determining the
nature of these sources, this FLAMINGOS-2 Galactic Center Survey will
have a dramatic impact on our knowledge of the Galactic accreting
binary population.

\vspace{-2mm}
\section{X-ray/IR Cross-Correlation}
\vspace{-1mm} 
We have cross-correlated the source catalog derived from the W02
\Chandra survey with \JHK images of 26 selected 2.5\arcsec$^{2}$
regions obtained with ISAAC on the VLT to identify candidate IR
counterparts to the X-ray sources \citep{reba}.  IR spectroscopy to
search for accretion signatures will be required for definitive
identifications.  Using this technique, we were able to conclusively
identify the IR counterpart to ``Edd-1'', one of these newly
discovered low-luminosity X-ray sources \citep{edd1}.

Cross-correlation of our CTIO ISPI \JHK image (Fig. 1b) of the central
10 pc$^{2}$ of the GC with the deep Sgr A* image of M03 is underway;
this will produce a large number of IR candidate counterparts to the
X-ray sources.  Due to the extremely high stellar density in the
Nuclear Bulge, many of these astrometric ``matches'' are likely to be
chance superpositions.  With thousands of candidate counterparts,
traditional long-slit single-target spectroscopy would be a
prohibitively inefficient method by which to identify true
counterparts.  Thus we will need to follow-up with multi-object IR
spectroscopy to find the true matches: this is the work which will be
performed with the FLAMINGOS-2 Galactic Center Survey (F2GCS).

\begin{figure}
\centering
\subfigure[\Chandra/ACIS-I] 
{
    \label{fig:sub:a}
    \scalebox{0.5}{\includegraphics{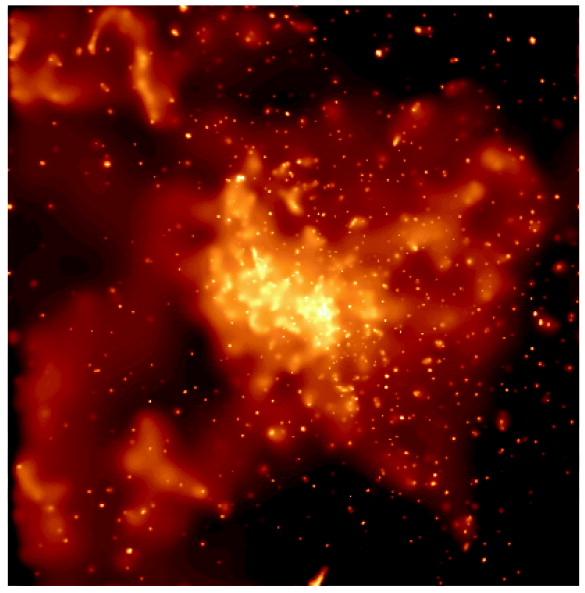}}
}
\hspace{1cm}
\subfigure[CTIO/ISPI] 
{
    \label{fig:sub:b}
    \scalebox{0.4}{\includegraphics{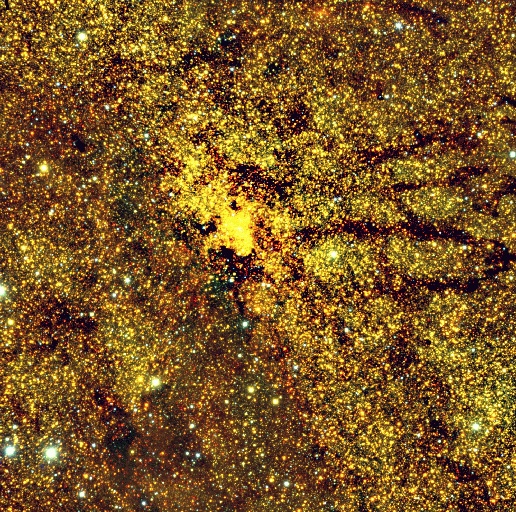}}
}
\caption{(a) \Chandra image of the central 8.5\arcmin$^{2}$ of the GC \citep{muno03}.  (b) Our CTIO/ISPI \JHK image of the same region.}
\label{fig:sub} 
\vspace{-3mm}
\end{figure}

\vspace{-2mm}
\section{Extinction}
\vspace{-1mm} 
We find that the IR extinction in the GC can vary on scales as small
as 5\arcsec~ (0.2-0.6 pc at 8 kpc; \cite{ajg06}).  Some areas show
little evidence of this ``granularity'', while others are highly
structured.  To obtain reddening-corrected stellar photometry, a local
value for the \JHK extinction (on scales $< 20\arcsec$) must be
measured and applied.  The relationship of extinction to wavelength in
the IR is a power law with slope $\alpha$ \citep{mw90}.  The
``canonical'' value for $\alpha$ is $\sim$2 (\eg \cite{rl85,nish}).
In contrast, for the GC we find a mean value of $\alpha =
2.64\pm0.52$; and furthermore, along any given line of sight to the GC
$\alpha$ varies substantially, ranging from $\sim$1.8--3.6
\citep{ajg08}.  Thus we find that the ``universal'' IR extinction law
is {\it not} universal in the GC!

\vspace{-2mm}
\section{Combining UKIDSS and \Chandra}
\vspace{-1mm} 

\begin{figure}
\centering
\subfigure[UKIDSS GPS mosaic of the GC] 
{
    \label{fig:sub:a}
    \scalebox{0.2}{\includegraphics{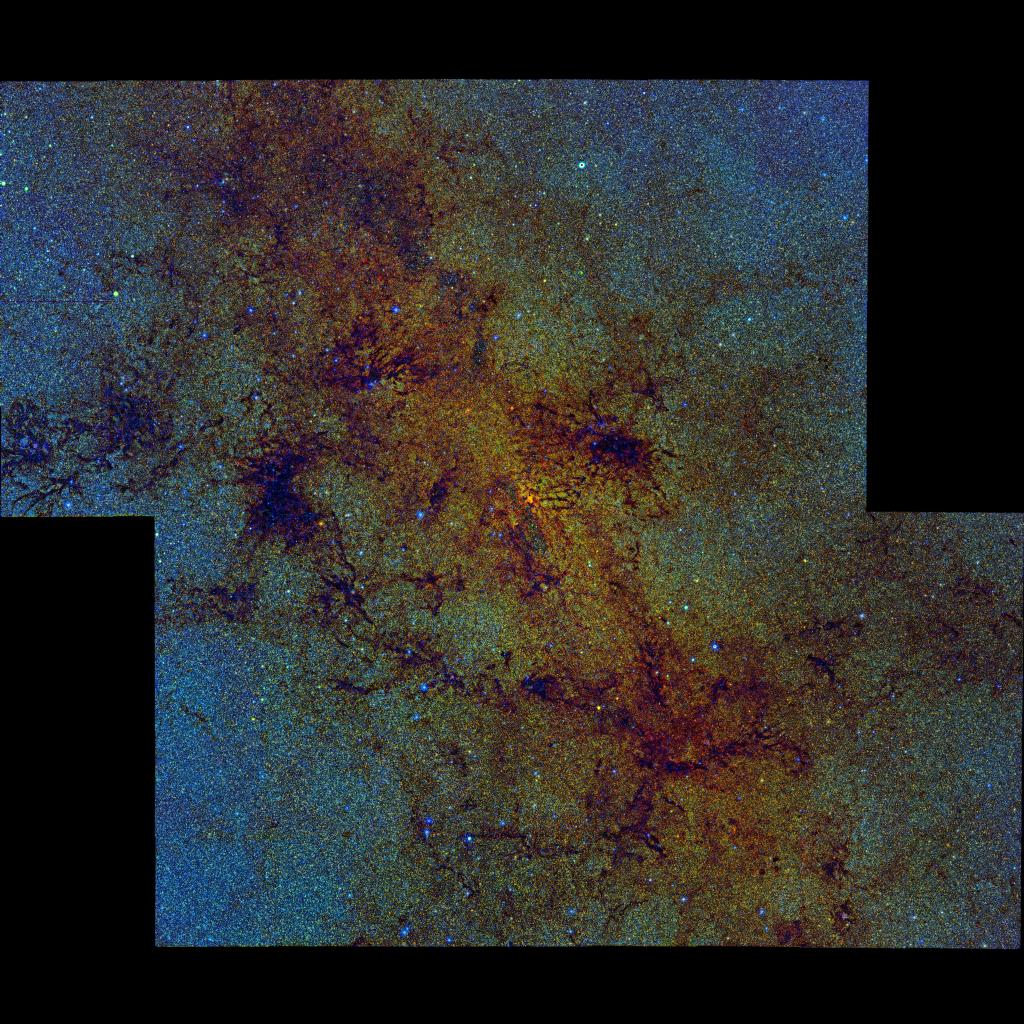}}
}
\hspace{1cm}
\subfigure[UKIDSS extinction map of the GC] 
{
    \label{fig:sub:b}
    \scalebox{0.27}{\includegraphics{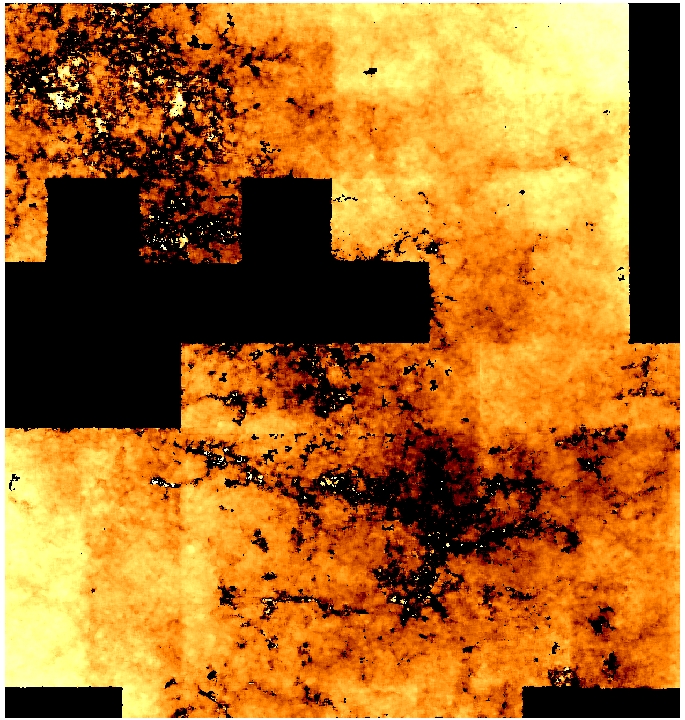}}
}
\caption{(a) UKIDSS GPS \JHK mosaic of the GC which covers the entire W02 survey area \citep{gps}.  (b) Extinction map of the same area of the GC, derived from the GPS data; black areas indicate regions which have not yet been fully included in the UKIDSS second data release (Gosling \etal, {\it these proceedings}).}
\label{fig:sub} 
\vspace{-3mm}
\end{figure}

The United Kingdom Infrared Deep Sky Survey (UKIDSS) is an imaging
survey covering 7500\degs$^{2}$ of the northern sky using a wide-field
IR camera (WFCAM) on UKIRT.  The Galactic Plane Survey (GPS) component
of UKIDSS covers the Nuclear Bulge (-2\degs$<l<$15\degs, $|b|<$2\degs)
to a depth of \JHK $\sim$18 with a resolution of $\sim$0.8\arcsec~
\citep{gps}.  We have obtained an additional 600 ksec of \Chandra
observations of the central degree of the GC; cross-correlation of the
resultant X-ray source catalog, combined with the earlier W02 and M03
datasets, with the UKIDSS GPS imaging of the GC (Fig. 2a) will provide
additional targets for the F2GCS. With the UKIDSS GPS data we can also
map the 1-2 $\mu$m ``granularity'' in the GC (Fig. 2b) and compare it
to the Spitzer map of the same region \citep{spitzer}, to search for
correlations in the dust and gas structure.


\end{document}